\documentclass[twocolumn,showpacs,preprintnumbers,amsmath,amssymb,aps]{revtex4}

\usepackage{graphicx}
\usepackage{dcolumn}
\usepackage{bm}
\usepackage{hyperref}

\begin{document}


\title{Single-photon two-qubit SWAP gate for entanglement manipulation}

\author{Marco Fiorentino} \email{mfiore@mit.edu}
\author{Taehyun Kim}
\author{Franco N.\ C.\ Wong}
\affiliation{Research Laboratory of Electronics, Massachusetts
Institute of Technology, Cambridge, MA 02139}

\begin{abstract}
A SWAP operation between different types of qubits of single
photons is essential for manipulating hyperentangled photons for a
variety of applications. We have implemented an efficient SWAP
gate for the momentum and polarization degrees of freedom of
single photons.  The SWAP gate was utilized in a single-photon
two-qubit quantum logic circuit to deterministically transfer
momentum entanglement between a pair of down-converted photons to
polarization entanglement. The polarization entanglement thus
obtained violates Bell's inequality by more than 150 standard
deviations.
\end{abstract}

\pacs{03.67.-a, 03.67.Lx, 42.50.Dv, 03.67.Mn}

\maketitle

Linear optical quantum computation (LOQC) has recently attracted
great interests following the demonstration \cite{KLM} that a
scalable quantum computer based on linear optical components is
possible. It has also been known that linear optical systems could
achieve non-scalable quantum computation by encoding multiple
qubits in several degrees of freedom of a single photon
\cite{detth}. Experiments in the latter were limited to a few
qubits due to the complexity of the optical setup \cite{detex} and
they did not use entanglement resources. Recently, however,
several groups have proposed the use of deterministic logic gates
in conjunction with sources of entangled or hyperentangled (i.e.,
entangled in more than one degree of freedom) photons to execute
simple quantum protocols. The combination of deterministic logic
and entangled photons can be used for one-shot demonstration of
nonlocality with two observers \cite{Zeilinger}, complete
measurement of Bell's states \cite{Monken}, cryptographic
protocols \cite{genovese}, and quantum games \cite{HPgame}. These
proposed experiments rely on the ability to create hyperentangled
states and successively project them onto suitable sets of basis
states for measurement. Manipulation of entanglement would benefit
significantly from efficient deterministic one- and two-qubit
gates thus permitting hyperentangled photons to be used as
essential quantum resources.

In the case of hyperentanglement in the polarization and momentum
(spatial) degrees of freedom of a single photon, single-qubit
rotation can be accomplished using wave plates and beam splitters.
We have recently demonstrated a single-photon two-qubit (SPTQ)
implementation of a deterministic controlled NOT (CNOT) gate that
operates on the momentum and polarization degrees of freedom of
single photons \cite{us}. It is well known that any arbitrary
unitary operation can be generated using CNOT gates and
single-qubit rotations, which can be used to manipulate qubits of
single or entangled photons.  In this letter we apply SPTQ logic
to manipulate entanglement between two photons. Specifically, we
have built a SWAP gate and transferred the entanglement in the
momentum degree of freedom of a pair of down-converted photons to
their polarization. This type of transfer is fundamentally
different from ``entanglement swapping" as described in
Ref.~\cite{swapping}: our SWAP operation involves two different
qubits of the same photon, whereas conventional entanglement
swapping is between the same type of qubit of two different
photons. Our experiment is both the first application of SPTQ
logic to entangled photons and a verification of the momentum
entanglement of down-converted photons. Compared to similar
proposals \cite{Zeilinger,Monken} our implementation of SPTQ logic
has the advantage of relying on gates that are robust and require
no active path length stabilization, therefore simplifying the
optical layout. The ability to swap two qubits constitutes an
important step toward the realization of proposed SPTQ protocols
\cite{Monken,Zeilinger,genovese,HPgame}. For example, some
single-qubit operations necessary to implement these protocols,
such as single-qubit rotations and projections onto the
$(|0\rangle +|1\rangle, |0\rangle - |1\rangle)$ basis, require
phase-stable interferometers for the momentum qubit.  With the
SWAP gate, one can implement these operations in the polarization
domain simply with wave plates and polarizers.

\begin{figure}[t]
\centerline{\rotatebox{0}{\scalebox{0.5}{\includegraphics{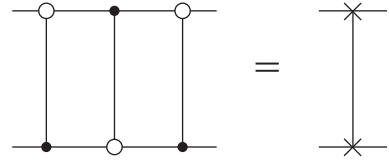}}}}
\caption{Schematic of the SWAP gate logic circuit.} \label{Logic}
\end{figure}

For the quantum resource in our experiment we exploit the
intrinsic momentum entanglement of down-converted photon pairs.
This type of entanglement has been demonstrated by Rarity and
Tapster \cite{ratap} and is based on the conservation of momentum
in the parametric down-conversion process. The state of the
down-conversion output can be derived from Eq.\ 7 in Ref.\
\cite{mete}. For simplicity we assume the pump to be a
monochromatic plane wave propagating along the crystal's principal
$x$ axis. The state is given by
\begin{eqnarray}
|\Psi\rangle_{IN} & \simeq & \int d\textbf{q}_S d\omega_S L
e^{-i\frac{L \Delta}{2}}  \textrm{sinc}\left(\frac{L \Delta}
{2}\right) \\ \nonumber & &\hat{a}_H^\dagger\left(\textbf{q}_S,
\omega_S \right) \hat{a}_V^\dagger\left(-\textbf{q}_S, \omega_P -
\omega_S \right) |0\rangle, \label{prestate}
\end{eqnarray}
where the integral is a triple integral that extends to the whole
plane spanned by the transverse (with respect to $x$) component
$\textbf{q}_S$ of the signal wavevector and over the range of
positive frequencies spanned by the signal frequency $\omega_S$.
The creation operators refer to the horizontally ($H$) and
vertically ($V$) polarized signal and idler, respectively. $L$ is
the crystal length, $\omega_P$ is the pump frequency, and $\Delta$
is the phase mismatch as defined in Ref. \cite{mete}. Equation
(\ref{prestate}) shows the correlation in momentum between signal
and idler photons. We now restrict our attention to two
propagation directions: one on the top $\textbf{q}_T$ and its
conjugate at the bottom $\textbf{q}_B =-\textbf{q}_T$. We take the
signal frequency to be $\omega_S= \omega_P/2$ and assume the phase
mismatch $\Delta$ to be zero. In the experimental setup the single
frequency and single direction constraints were enforced by the
use of interference filters and irises. The state then becomes
\begin{eqnarray}
|\Psi\rangle_{IN} & \simeq &
\left(\hat{a}_H^\dagger\left(\textbf{q}_T,\omega_P/2\right)
\hat{a}_V^\dagger\left(\textbf{q}_B,\omega_P/2\right) + \right. \\
\nonumber &+& \left.
\hat{a}_H^\dagger\left(\textbf{q}_B,\omega_P/2\right)
\hat{a}_V^\dagger\left(\textbf{q}_T,\omega_P/2\right) \right)
|0\rangle. \label{prestate2}
\end{eqnarray}
Equation (2) describes two photons that can be in four orthogonal
states: horizontally polarized top ($HT$), vertically polarized
top ($VT$), horizontally polarized bottom ($HB$), and vertically
polarized bottom ($VB$). Each photon is therefore described by a
state in a four dimensional Hilbert space. Following Ref.\
\cite{Zeilinger} we rewrite each four-dimensional Hilbert space as
the tensor product of two two-dimensional Hilbert spaces (i.e.
qubits). In this formalism the normalized state (2) can be
rewritten as
\begin{eqnarray}
|\Psi\rangle_{IN} & = & \frac{1}{\sqrt{2}}(|T_S B_I\rangle + |B_S
T_I\rangle)\otimes |H_S V_I\rangle \\ \nonumber & \equiv &
\frac{1}{\sqrt{2}}(|0_{MS}1_{MI}\rangle +
|1_{MS}0_{MI}\rangle)\otimes |0_{PS}1_{PI}\rangle \,,\label{state}
\end{eqnarray}
In the final expression we identify the $H$ and $T$ states with
the logical 0 and the $V$ and $B$ states with the logical 1 for
the four qubits designated as polarization ($P$) and momentum
($M$) of the signal ($S$) and idler ($I$). From Eq.~\ref{state} it
is clear that the photons emitted by the crystal are not
polarization entangled in general, unless signal and idler photons
are indistinguishable spectrally (frequency degenerate) and
temporally (timing compensated) \cite{twos,shih}, in which case
the $T$ and $B$ beams are polarization entangled, as demonstrated
in Ref.~\cite{Kwiat}. In the present experiment we ensure that the
photons are not polarization entangled by not compensating the
birefringence-induced time delay.

Manipulation of the four-qubit state of Eq.~\ref{state}, two for
each photon, can be achieved using SPTQ logic. We have previously
demonstrated a high fidelity polarization-controlled NOT (P-CNOT)
gate for SPTQ logic \cite{us} by use of a polarization Sagnac
interferometer with an embedded dove prism that flips and rotates
the input beam by 90$^\circ$. A momentum-controlled NOT (M-CNOT)
gate can be realized with a half-wave plate (HWP) oriented at
45$^\circ$ relative to the horizontal position and inserted in the
path of the $B$ beam. The SWAP we present here is a more complex
quantum gate that can be obtained by applying three consecutive
CNOT gates \cite{nielsen} as shown in Fig.~\ref{Logic}. A SWAP
gate exchanges the values of two arbitrary qubits without the need
of measuring them. For example, when applied to the arbitrary
two-qubit \emph{product} state $(\alpha |T\rangle +\beta
|B\rangle) \otimes (\gamma |H\rangle +\delta |V\rangle)$ a SWAP
gate transforms it into the state $(\gamma |T\rangle +\delta
|B\rangle) \otimes (\alpha |H\rangle +\beta |V\rangle)$. Note that
a SWAP acting on a qubit that is part of an entangled pair of
qubits transfers the entanglement to the other qubit, which may be
more conveniently manipulated. In the case of hyperentangled
photons, for example, swapping the entanglement from the momentum
to the polarization qubit allows a complete and unequivocal proof
of the successful generation of hyperentanglement. In our logic
protocol applying a sequence of a M-CNOT followed by a P-CNOT and
another M-CNOT realizes a SWAP gate. A SWAP gate applied to both
photons in the initial state $|\Psi\rangle_{IN}$ yields the
polarization-entangled state
\begin{eqnarray}
|\Psi\rangle_{OUT} = \frac{1}{\sqrt{2}} |0_{MS}1_{MI}\rangle
\otimes (|0_{PS}1_{PI}\rangle +
|1_{PS}0_{PI}\rangle)\,.\label{eq1}
\end{eqnarray}
Observe that if we omit the last M-CNOT gate the output state is
\begin{eqnarray}
|\Psi'\rangle_{OUT} = \frac{1}{\sqrt{2}} |0_{MS}1_{MI}\rangle
\otimes (|0_{PS}0_{PI}\rangle +
|1_{PS}1_{PI}\rangle)\,.\label{eq2}
\end{eqnarray}
which is also polarization entangled. Signal and idler photons in
$|\Psi\rangle_{OUT}$ and $|\Psi'\rangle_{OUT}$ are in a definite
momentum state (signal and idler are on opposite sides). Therefore
they can be separated with a mirror that reflects one part of the
beam and not the other. It is worth noticing that the entanglement
swapping presented here is deterministic, i.e., in principle all
the momentum-entangled photon pairs are converted into
polarization-entangled pairs.

\begin{figure}[tb]
\centerline{\rotatebox{0}{\scalebox{0.37}{\includegraphics{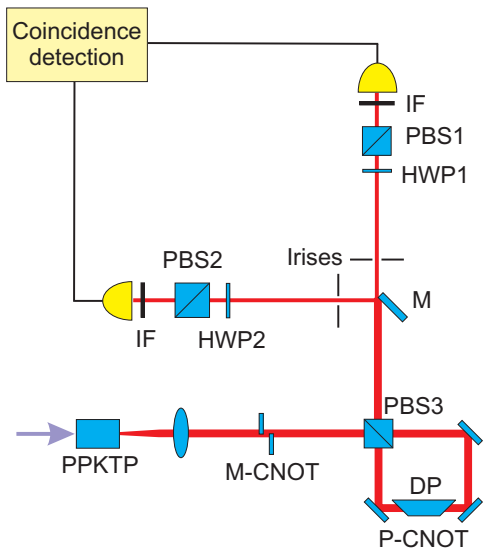}}}}
 \caption{Schematic of experimental setup. PPKTP: periodically
poled KTP crystal. PBS: polarizing beam splitter. DP: dove prism.
HWP: half-wave plate. IF: 1-nm interference filter. M: mirror.
P-CNOT: polarization-controlled NOT gate. M-CNOT: momentum
controlled NOT gate.} \label{Setup}
\end{figure}

Figure \ref{Setup} shows our experimental setup. We used pairs of
down-converted photons from a 1-cm-long periodically poled
potassium titanyl phosphate (PPKTP) crystal that was
continuous-wave pumped at 398.5 nm for type-II phase-matched
frequency-degenerate parametric down-conversion \cite{twos}. The
crystal temperature was adjusted so that signal and idler photons
were emitted in two overlapping cones with an external full
divergence of $\sim$13 mrad. The momentum modes were chosen with
two apertures after the gates instead of a two-hole aperture mask
placed before the gate as was done in Ref.~\cite{us}. We observed
a higher gate fidelity with the separate apertures after the
gates, probably due to slight size mismatch of the two-hole mask.
In our experimental realization of entanglement swapping we used
the same physical gates to manipulate both photons of the pair.
The two photons crossed the gates at different times owing to the
delay accumulated in the PPKTP crystal and therefore no
interference between them took place. The M-CNOT gate was a HWP
cut in a half-circular D shape with the fast axis forming a
45$^\circ$ angle with the $H$ direction. The plate was aligned so
that it only affected the bottom section of the beam. A second
HWP, identical to the first except for the fact that its axis was
parallel to the $H$ axis, was put in the path of the top part of
the beam to compensate for the time delay introduced by the first
HWP. The compensating wave plate was slightly tilted to obtain
optimal visibility in the entangled state analysis. This tilting
changed the length of the top beam path thus allowing one to
correct for path mismatch. The P-CNOT gate was a polarization
Sagnac interferometer with an embedded dove prism \cite{us}. The
input polarizing beam splitter (PBS3) of the P-CNOT gate directed
horizontally (vertically) polarized input light to travel in a
clockwise (counterclockwise) direction. As viewed by each beam,
the dove prism orientation was different for the two
counter-propagating beams. Therefore the top-bottom ($T$--$B$)
sections of the input beam were mapped onto the right-left
($R$--$L$) sections of the output beam for $H$-polarized light but
onto the $L$-$R$ sections for $V$-polarized light. If we identify
$|H\rangle$, $|T\rangle$, and $|R\rangle$ with the logical
$|0\rangle$ and $|V\rangle$, $|B\rangle$, and $|L\rangle$ with the
logical $|1\rangle$ it is easy to recognize that this setup
implements a CNOT gate in which the polarization is the control
qubit and the momentum (or spatial) mode is the target qubit.
After the P-CNOT gate the state of the photon pair is described by
Eq.~\ref{eq2}; we separated signal and idler photons using the
mirror M shown in Fig.~\ref{Setup} that reflected only the right
section of the beam. Signal and idler beams were then separately
sent through a 2.2-mm iris, a polarization analyzer formed by a
HWP and a polarizer, and a 1-nm interference filter centered at
797 nm. Besides being used for polarization analysis, wave plate
HWP2 in Fig.~\ref{Setup} assumed the role of the second M-CNOT
gate, thus completing the SWAP circuit.  The photons were detected
with single-photon counting modules (PerkinElmer SPCM-AQR-14) and
we measured signal-idler coincidences through a fast AND gate with
a 1-ns coincidence window \cite{Taehyun}. Given the short
coincidence window and the observed count rates (singles rates
$\leq 100,000$ counts/s), accidental coincidences were negligible.

\begin{figure}[tbh]
\centerline{\rotatebox{270}{\scalebox{0.4}{\includegraphics{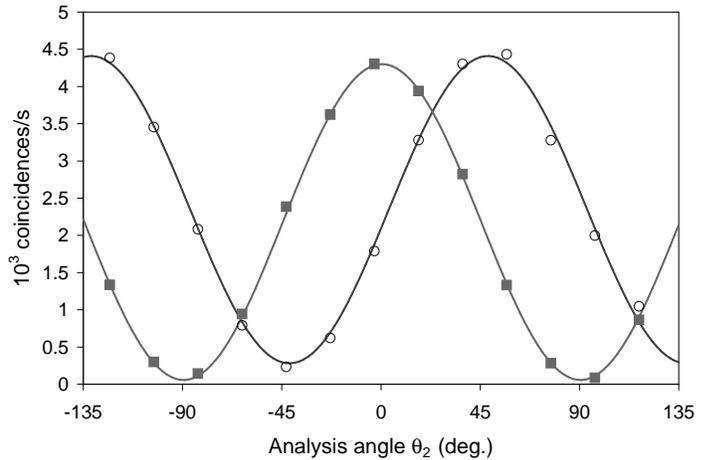}}}}
 \caption{Coincidence rates as a function of the polarization analysis angle
$\theta_2$ in arm 2 when the analyzer in arm 1 was set at an angle
$\theta_1 = 0^\circ$ (solid squares) and $45^\circ$ (open
circles). The lines are sinusoidal fits to the data.}
\label{Results}

\end{figure}

To test the performance of the SWAP gate we analyzed the resultant
polarization entanglement. Figure \ref{Results} shows the
coincidence rates versus the polarization analysis angle
$\theta_2$ in arm 2 of Fig.~\ref{Setup} when the analyzer in arm 1
was set at $0^\circ$ (solid squares) and $45^\circ$ (open
circles). The visibility of the sinusoidal fits is $V_0=97 \pm 2
\%$ for $0^\circ$ data and $V_{45}= 88 \pm 2 \%$ for the
$45^\circ$ data. The difference in visibility is due to the fact
that $V_{45}$ is more sensitive than $V_{0}$ to the imperfections
of the source and the gate interferometer. A measurement of the S
parameter \cite{Aspect} for the Clauser-Horne-Shimony-Holt form of
the Bell's inequality gives a value of $2.653 \pm 0.004$ that
violates the classical limit of 2 by more than 150 standard
deviations. These results clearly show that our SWAP gate had a
good fidelity and that the down-converted photons were indeed
initially momentum entangled.

The $V_{45}$ results in Fig.~\ref{Results} include errors due to
imperfect interference at the gate (gate fidelity) and incomplete
momentum entanglement of the source (source fidelity). To
determine how well our setup approximates the ideal SWAP gate it
is useful to separate the two contributions.  As a test we sent an
attenuated laser beam (filtered through a single-mode fiber and
collimated with an aspheric lens) through the gate, with the laser
frequency being the same as that of the down-converted photons. By
injecting the laser with a linear polarization oriented at
45$^\circ$ relative to the $H$ direction we measured the classical
visibility of the SWAP gate. This test measurement gave a
visibility $V_{C1}$ of $\sim$93\% for the gate. We also verified
that the M-CNOT gate did not affect the classical visibility in a
measurable way. The classical measurement was repeated without the
dove prism in  the polarization Sagnac interferometer (of the
P-CNOT gate) that yielded a visibility $V_{C2}$ of $\sim$95\%. The
2\% difference in the classical visibility ($V_{C1}-V_{C2}$) can
be attributed to either imperfections in the dove prism or
asymmetries in the injected laser beam. To further evaluate the
cause, we repeated the test experiment with a polarization Sagnac
interferometer in a triangular configuration that was insensitive
to input beam asymmetry.  In this configuration the interference
at the $T$ position at the output originated from the same spot of
the injected beam for both polarizations,  with or without the
dove prism (and similarly for the $B$ position at the output). For
the triangular configuration we obtained a difference in classical
visibility with and without the dove prism of $\sim$2.5\% that is
comparable to that of the non-triangular configuration, indicating
that the dove prism was responsible for a loss of $\sim$2\% in the
visibility of the P-CNOT gate.  The remaining $\simeq$5\% loss of
classical visibility $V_{C1}$ can be attributed to wavefront
distortions introduced by the beam splitter cube (which leads to
our continuing effort to obtain  a polarizing beam splitter with a
low wavefront distortion in both transmission and reflection).
Given our quantum visibility $V_{45}$ of 88\% and the classical
test measurement results of the P-CNOT interferometer we conclude
that the source fidelity was 95\% that was limited by
imperfections in the momentum entanglement of the down-conversion
source (probable causes: defects in PPKTP crystal poling and
wavefront distortions of the downconverted beams).  The SWAP gate
fidelity was 93\% and was limited by less than ideal components
(polarizing beam splitter and dove prism).

In conclusion we have demonstrated a high fidelity SWAP gate for
single-photon two-qubit logic. To realize such a gate we have
built an essential set of gates in the SPTQ quantum logic family
comprising linear optical P-CNOT and M-CNOT gates that are robust
and do not need active length stabilization. We applied the SWAP
gate to momentum-entangled photons thereby transferring the
entanglement from the momentum to the polarization degree of
freedom. This is, to the best of our knowledge, the first
application of SPTQ linear optical quantum logic to entangled
photons. Our experiment opens the way to the demonstration of more
complex SPTQ manipulation of entanglement including the
manipulation of 3- and 4-photon states. This type of few-qubit
quantum information processing is at the core of a number of
applications ranging from single-shot two-observers demonstration
of nonlocality \cite{Zeilinger} to two-qubit quantum key
distribution \cite{genovese}.

This work was supported by the the DoD Multidisciplinary
University Research Initiative (MURI) program administered by the
Army Research Office under Grant DAAD-19-00-1-0177.

\end{document}